\documentstyle[referee]{l-aa}
\begin{document}

   \thesaurus{11.17.3; % quasars: general
              11.07.1} % gravitational lensing

   \title{A remark on the inversion of the magnification bias 
       in the quasar-galaxy associations}

   \author{Zong-Hong Zhu\inst{1} and Xiang-Ping Wu\inst{2} 
%          \inst{1}
          }

   \offprints{Z. H. Zhu}

   \institute{$^1$Department of Astronomy, Beijing Normal University, 
       Beijing 100875, China\\
     $^2$Beijing Astronomical Observatory, Chinese Academy of Sciences,
       Beijing 100080, China
             }

   \date{Received 00 00, 1996; accepted 00 00, 1997}

   \maketitle
   \markboth{Zhu, Wu: Inversion of magnification bias}{ }

   \begin{abstract}

We present an alternative but intuitive method of obtaining 
the true or intrinsic quasar luminosity function through the 
observed one in the framework of gravitational lensing. 
Such a so-called inversion of the magnification bias   
can be straightforwardly  performed by means of the Mellin transformation 
instead of other complex methods. Application of our approach to the 
well-known form of the observed quasar luminosity function identifies the
previous results in literature.


   \end{abstract}

%
%  14.Sep.'90: Demo-Vs.
%________________________________________________________________

It has been argued for decade whether the magnification bias 
due to gravitational lensing by the matter clumps in the universe 
affects our quasar number counts $N_o(S)$, or equivalently the 
determination of quasar luminosity function $\Phi_o(L,z)$, 
and therefore, accounts at least partially for
the apparent evolution of quasars (Turner 1980;1981; Canizares 1981;1982;
Avni 1981; Vietri \& Ostriker 1983; Schneider 1986;1992; 
Kayer \& Refsdal 1988; Pei 1995). In particular, an overdensity of background 
quasars near foreground galaxies, clusters and even quasars would occur
as a result of the magnification bias [see Wu (1996) for a recent summary].
All these issues can be simplied as a convolution of the true
quasar number count $N_t(S)$
or luminosity function $\Phi_t(L,z)$ around redshift $z$ with
a magnification probability function $p(\mu)$ or $p(\mu|z)$:
%1
\begin{equation}
N_o(S) = \int_0^{\infty} d\mu p(\mu) \mu^{-1} N_t(\mu^{-1}S),
\end{equation}
or
%2
\begin{equation}
\Phi_o(L,z) = \int_0^{\infty} d\mu p(\mu|z) \mu^{-1} \Phi_t(\mu^{-1}L,z),
\end{equation}
where $S$ is the flux threshold of the quasar sample, $L$, the absolute
luminosity and, $\mu$, the magnification factor. It is evident that 
the magnification probability distribution should satisfy the following 
constraints
%3,4
\begin{eqnarray}
        \int_0^{\infty} d\mu p(\mu|z) =1,\\
	\int_0^{\infty} d\mu \mu p(\mu|z)=1,
\end{eqnarray}
which correspond to, respectively, the normalization and the flux 
conservation. The question now reduces to how to find the true quasar
number  count $N_t(S)$ or luminosity function $\Phi_t(L,z)$
using the observed quantities for a given magnification probability
distribution $p(\mu)$ or $p(\mu|z)$, namely, the inversion of
eq.(1) or eq.(2).   To solve eq.(1), Schneider (1992) utilized the 
Volterra equation of the second kind with a kernel function
$K(S,x)$ combined with other mathematical techniques and
a specific boundary value, whereas
for eq.(2) Pei (1995) performed an expansion of the true 
luminosity function $\Phi_t(L,z)$ into the Taylor series coupled with
a symbolic operator method to separate the variable $\mu$ from 
$\Phi_t(L,z)$.  These sophisticated methods should  be in principle
applicable to various matter distributions, 
allowing us to derive the true quasar number count or luminosity function. 
However, the actual application of these methods often turns to be
complicated, aside from the unknown details of the
magnification probability function.   Motivated by the importance of
eq.(1) or eq.(2) in the study of the associations of angular positions of
distant quasars with foreground objects (galaxies, groups and clusters of 
galaxies and quasars), we present an alternative but intuitive  
approach to the inversion of eq.(2). Similarly, this approach 
can be equivalently employed for the inversion of eq.(1).

Our method is based on the Mellin transformation (Titchmarsh 1948). 
For a give function of $f(x)$, its Mellin transformation is 
an integral of
%5
\begin{equation}
	\tilde{f}(s) = \int_0^{\infty} x^{s-1} f(x) dx
\end{equation}
so that the source function itself reads
%6
\begin{equation}
	f(x) = \frac{1}{2\pi i} \int_{c-i\infty}^{c+i\infty}
               x^{-s} \tilde{f}(s) ds,
\end{equation}
where $c$ should be properly chosen to ensure that the singularities of 
the function $x^{-s}\tilde{f}(s)$ are on the left of the routine.
The Mellin transformation of eq.(2) is thus
%7
\begin{equation}
\tilde{\Phi}_o(s,z) = \int_0^{\infty}\mu^{s-1}p(\mu|z)d\mu
                      \int_0^{\infty}(\mu^{-1}L)^{s-1}
                       \Phi_t(\mu^{-1} L,z)d(\mu^{-1}L),
\end{equation}
in which we have adopted a physically reasonable boundary  that 
the magnification probability
function $p(\mu|z)\longrightarrow 0$ for a sufficiently large
magnification $\mu$.   
While the right-hand side of the above equation is fortunately
a product of two Mellin transformations of functions 
$p(\mu)$ and $\Phi_t(\mu^{-1}L,z)$, we have
%8
\begin{equation}
\tilde{\Phi}_o(s,z) = \tilde{p}(s|z) \cdot \tilde{\Phi}_t(s,z),
\end{equation}
where $\tilde{p}(s|z)$ is the Mellin transformation of $p(\mu|z)$
and $\tilde{\Phi}_t(s,z)$ and $\tilde{\Phi}_o(s,z)$ are the corresponding
quantities of $\Phi_t(L,z)$ and $\Phi_o(L,z)$, respectively. From 
eqs.(6) and (8) the true quasar luminosity function can be expressed as
%9
\begin{equation}
	\Phi_t(L,z) = \frac{1}{2\pi i} \int_{c-i\infty}^{c+i\infty}
                      L^{-s} \tilde{\Phi}_t(s,z) ds.
\end{equation}
As it has been shown, our procedure of inversion of the magnification
bias is more straightforward and convenient in application than 
the previous methods.

To demonstrate how efficiently the present method works, 
we take the exponential $L^{1/4}$ form of the
observed quasar luminosity function used by Pei (1995)
%10
\begin{equation}
        \Phi_o (L,z) = \frac{\Phi_*}{L_z} 
             {\left(\frac{L}{L_z}\right)}^{-\beta} 
             \exp\left[-{\left(\frac{L}{L_z}\right)}^{1/4}\right]
\end{equation}
with the luminosity evolution 
%11
\begin{equation}
	L_z = L_* (1+z)^{-(1+\alpha)} \exp [-(z-z_*)^2/2\sigma_*^2],
\end{equation}
here $(\Phi_*, \beta, \alpha, z_*, \sigma_*)$ are the parameters fitted by
observations. Applying the Mellin transformation of eq.(5) to eq.(10)
yields
%12
\begin{equation}
	\tilde{\Phi}_o(s,z) = 4 \frac{\Phi_*}{L_z} {L_z}^s 
                              \Gamma\left[4(s-\beta)\right],
\end{equation}
where $\Gamma$ is the usual $\Gamma$-function with a variable of
$4(s-\beta)$. If we adopt the same notation as Pei (1995) by defining
%13
\begin{equation}
	Z(s-1|z) \equiv \ln\langle \mu^{s-1} \rangle = 
        \ln\left[ \int_0^{\infty} d\mu \mu^{s-1} p(\mu|z)\right]
\end{equation}
then the Mellin transformation of $p(\mu|z)$ reads 
%14
\begin{equation}
	\tilde{p}(s|z) = \exp\left[ Z(s-1|z)\right]
\end{equation}
A combination of eq.(12) and eq.(14) gives rise to 
the Mellin transformation of the true quasar luminosity function
$\tilde{\Phi}_t(s,z)=\tilde{\Phi}_o(s,z) [\tilde{p}(s|z)]^{-1}$, and
the inverse Mellin transformation of $\tilde{\Phi}_t(s,z)$ 
results in the true quasar luminosity function 
%15
\begin{equation}
   \Phi_t(L,z) = \frac{1}{2\pi i} 
                 \int_{(\beta+1)-i\infty}^{(\beta+1)+i\infty} 4 
                 \frac{\Phi_*}{L_z} {\left(\frac{L}{L_z}\right)}^{-s} 
     \Gamma\left[4(s-\beta)\right] \cdot \exp\left[-Z(s-1|z)\right] ds,
\end{equation}
in which we have chosen $c=\beta+1$. Finally, the ratio of the true
quasar luminosity function to the observed one is simply
%16
\begin{equation}
     \frac{\Phi_t}{\Phi_o} = \exp(-4m+e^m) \times 
                   \int_{-\infty}^{+\infty} ds \; \Gamma(4+2\pi is) 
            \cdot \exp\left[-2\pi ism - Z(\beta_r+2\pi is\beta_i|z)\right],
\end{equation}
i.e., the result of Pei (1995) [eq.(39)], 
where $m \equiv \beta_i \ln(L/L_z)$, $\beta_r \equiv \beta$, 
and $\beta_i \equiv 1/4$. 

Nevertheless, we point out that 
a quantitative analysis of  $\Phi_t/\Phi_o$ depends on  
the observed quasar luminosity function $\Phi_o(L,z)$ and 
the magnification probability function $p(\mu|z)$. Except for some
specific forms of  $\Phi_o(L,z)$ [e.g. eq.(10)] and $p(\mu|z)$, numerical 
computations should be often employed in finding the Mellin transformations 
$\tilde{\Phi}_o(s,z)$ and $\tilde{p}(s|z)$, and hence, the true quasar
luminosity function $\Phi_o (L,z)$. In particular, it is relatively hard to
get a simple form of $p(\mu|z)$ when the lens exhibits a complicated
matter structure (see, for example, Schneider 1992).  
A detailed investigation for various objects as lenses
is beyond the scope of this short note. 
We emphasize that the present method may 
be useful in the study of the associations of background quasars
with foreground objects. Recall that the association problems, if real, 
have not been well account for to date in terms of gravitational lensing
(Zhu et al. 1996). One of the possibilities is to abandon 
the unaffected background hypothesis, namely, the observed quasar
number-magnitude relation or luminosity function 
has probably been contaminated by
gravitational lensing according to eq.(1) or eq.(2). 
A further study based on more realistic lensing models,  
incorporating with the cosmological simulations of formation and
evolution of large-scale structures, would provide a helpful insight 
into the problem.

\begin{acknowledgements}
Valuable comments by the referee, Peter Schneider, are gratefully 
acknowledged. This work was supported by the National Science Foundation
of China.
\end{acknowledgements}

\end{document}